\documentclass[12pt]{article}

\usepackage{amssymb,amstext,amsmath,amsthm}
\usepackage[dvips]{graphicx}
\usepackage{latexsym}
\usepackage{psfrag}
\usepackage{amsfonts}
\usepackage{bbm}

\newcommand{\Ref}[1]{(\ref{#1})}

\newcommand{\eqa}{\begin{eqnarray}}
\newcommand{\neqa}{\end{eqnarray}}
\newcommand{\equ}{\begin{equation}}
\newcommand{\nequ}{\end{equation}}
\newcommand{\no}{\nonumber\\}

\newcommand{\p}{\partial}

\def\d{\delta}
\def\f{\frac}

\def\wtl{\widetilde}

\usepackage{bbm}

\let\eps=\epsilon

\let\si=\sigma
\newcommand{\bd}{\mathbf d}

\begin{document}

\title{\Large\bf Linearized dynamics from the 4-simplex Regge action}
\author{Bianca Dittrich, Laurent Freidel and Simone Speziale\footnote{bdittrich@perimeterinstitute.ca,  lfreidel@perimeterinstitute.ca, sspeziale@perimeterinstitute.ca} 
\\ [1mm]
\em\small{Perimeter Institute, 31 Caroline St. N, Waterloo, ON N2L 2Y5, Canada.}}
\date{\small\today}

\maketitle

\begin{abstract}
We study the relation between the hessian matrix of the riemannian Regge action on a 4-simplex
and linearized quantum gravity. We give an explicit formula for the hessian as a function of the geometry,
and show that it has a single zero mode.
We then use a 3d lattice model to show that (i) the zero mode is a remnant of the continuum diffeomorphism invariance,
and (ii) we recover the complete free graviton propagator in the continuum limit.
The results help clarify the structure of the boundary state needed in the recent calculations
of the graviton propagator in loop quantum gravity, and in particular its role in fixing the gauge.
\end{abstract}

\section{Introduction}
Loop quantum gravity (LQG) and the spinfoam formalism \cite{carlo} are candidate background independent quantizations of general relativity (GR).
A key open question to check the consistency of this approach is whether it gives the right semiclassical limit.
Recently there have been important advances towards a positive answer 
in cosmological models \cite{cosmo}, black hole physics \cite{bh},
canonical LQG \cite{Giesel} and in the linearized expansion \cite{CarloLeo, grav}.
In particular in \cite{grav} it was studied if the riemannian theory admits a regime in which the low energy description of linearized quantum gravity emerges. The basic quantity of linearized quantum gravity is the 
graviton propagator. This quantity has been defined in \cite{CarloLeo} and the promising results of \cite{grav} show that the correct spacetime dependence of some of its components is indeed emerging in a well defined limit.
The key for this mechanism is the intermediate step of linearized  quantum Regge calculus.
Linearized quantum Regge calculus has been studied in the literature \cite{Rocek, Ruth3d}, and the correctness
of its continuum limit poses the spinfoam calculations on solid grounds.
Yet there are many open questions in the spinfoam procedure, such as 
extending the formalism to arbitrary backgrounds, dynamically computing
the boundary state, and understanding the gauge fixing procedure. 
In this paper we investigate linearized quantum
Regge calculus from a perspective that is close to the spirit of the spinfoam calculations, 
with the aim of shedding light on some of the open questions.

Our first result is a formula for the hessian of the riemannian 
Regge action on an arbitrary $n$-simplex configuration.
This formula generalizes the equilateral configuration case appeared in the literature
(see third reference in \cite{grav}), and it is
relevant for studying the graviton propagator, or 2-point function, around a generic background. 
The latter scales as the background distance between 
the two points to the power $2-n$, so one might naively expect a similar scaling for the hessian matrix.
The explicit formula we obtain shows that this is not the case, and furthermore shows that this matrix is not
invertible. This is reminiscent of the non-invertibility of the quadratic term of the 
linearized continuum theory due to diffeomorphism invariance, and raises the question whether the non-invertibility of the hessian can be traced back to the same symmetry.

To address both questions of scaling and non-invertibility, we consider a 3d infinite lattice model. 
In this model we show that adding up the hessian matrices of all the tetrahedra
in the full lattice one obtains the correct quadratic term of linearized Regge calculus. This in turn can be directly obtained discretizing the (quadratic term of the) continuum Einstein-Hilbert action.
Upon properly gauge fixing this quadratic term, 
we obtain the correct free propagator, including both its distance dependence
and its tensorial structure. This is our second result, and it allows us to answer the above questions.
First, the correct scaling is indeed encoded in the hessian matrix, and shows up explicitly when
a full lattice is considered. 
Second, the non-invertibility is the remnant of the diffeomorphisms on the lattice, and 
the full lattice gauge fixing term induces a gauge fixing term for a single hessian matrix.
In our analysis we benefited from the techniques and results of \cite{Ruth3d}, which we develop further.

Based upon these results, we then discuss in more details the gauge fixing procedure and the general boundary formalism.
The discussion allows us to make a few important points concerning the spinfoam calculations of \cite{grav}. 
In particular, we argue that in those calculations the gauge is fixed by the boundary state. This means that the free parameters entering it have to be fixed not only by the dynamics, but also by the choice of gauge.
We give a prescription for computing dynamically the quadratic order of the boundary state.

Our results support once more the relevance of
Regge calculus to study the semiclassical limit of spinfoams. This idea has been often advocated
\cite{Immirzi}, and dates back to the original Ponzano-Regge model \cite{Ponzano}. The link between
spinfoam kernels and Regge calculus has been deeply studied \cite{asympt}, and recently extended
in 3d to expectation values of geometry \cite{Hackett} and to the coupling with Yang-Mills theory \cite{Speziale}.

This paper is organized as follows. In section \ref{Sec2} we briefly review conventional
length Regge calculus in any dimension $n$, discuss the modification where the $n-2$-volumes of the hinges
are the fundamental variables, and provide the explicit formula for the hessian.
In particular, we show that this matrix has a single zero mode related to the vanishing of the Gram determinant
for a flat $n$-simplex.
In section \ref{Sec4} we introduce the 3d infinite lattice model, and show how the hessian matrix
leads to the correct free graviton propagator.
In section \ref{secgf} we discuss in more details the gauge fixing procedure, and how the hessian zero mode
is related to the continuum diffeomorphism invariance. In section \ref{SecGB} we discuss the general boundary
formalism, and we stress a few key properties that the boundary state has to satisfy to provide the correct
physics. Finally, in section \ref{SecConcl} we summarize our results and discuss their perspective on the spinfoam 
calculations.

\section{Hessian of the Regge action}\label{Sec2} 
Regge calculus \cite{Regge} is a discrete approximation to GR where the fundamental
variables are the lengths $\ell_e$ of the edges of a triangulated manifold representing spacetime,
denoted $\Delta+\p\Delta$ where $\p\Delta$ is the boundary triangulation.
In any dimension $n$, the curvature is distributional and concentrated on the 
$n-2$ simplices of the triangulation, called hinges (edges in 3d, triangles in 4d).
The action is 
\equ\label{Regge}
S_{\rm R}[l_{e}] = \sum_{h\in \Delta} V_{h} \, \eps_{h} (l_{e}) + S_{\p\Delta}[l_e],
\nequ
where $V_{h}$ is the $(n-2)$-volume of the hinge $h$ and 
the deficit angles $\eps_{h}$ represent the curvature. 
They are defined as $\eps_{h} = 2\pi - \sum_\si \theta_{h}^\sigma$, where $\theta_{h}^\si$ is  the dihedral angle
of the $n$-simplex $\si$ sharing the hinge $h$. This is a function of the edge lengths of the $n$-simplex
as described below.
The boundary term is $S_{\p\Delta}[l_e] = \sum_{h} V_{h} \sum_\si \left({\pi}- \theta^\si_{h}(l_e)\right)$, 
and using the definition of the deficit angles, one can rewrite the full action as a sum over 
$n$-simplices,
\equ\label{Regge1}
S_{\rm R}[l_e] = \sum_\si s_\si[l_e], \qquad 
s_\si[l_e] \equiv \sum_{{h}\in \si \cap \Delta} V_{h} \left( \f{2\pi}{N_{h}}- \theta^\si_{h}(l_e)\right)
+ \sum_{{h}\in \si\cap{\p\Delta}} V_{h} \left( \f{\pi}{N_{h}}- \theta^\si_{h}(l_e)\right),
\nequ
where $N_{h}$ is the number of $n$-simplices hinged on ${h}$.
The continuum limit of \Ref{Regge} gives the Einstein--Hilbert action.

The link between LQG and Regge calculus is particularly transparent in 3d: 
in a well defined limit, the spinfoam partition function
for LQG is dominated by exponentials of the Regge action \Ref{Regge} (for details see for instance \cite{carlo}).
In 4d on the other hand, the same limit seems related to a modification of \Ref{Regge} in which the fundamental
variables are the hinge areas $A_t$ \cite{carloarea}. 
Area Regge calculus is much less known in the literature, and the key open question
is whether one can do the change of variables from lengths to areas in conventional Regge calculus. 
A positive answer would support area Regge calculus as an equally viable discrete approximation
of GR. This question has been only partially addressed so far, and at the present state of investigation,
two results are clear:

\begin{itemize}
\item A single 4-simplex has ten edges and ten triangles, 
thus one can directly study the Jacobian matrix for the change of variables from
lengths to areas, $\f{\p A_t}{\p l_e}$. The latter turns out to be 
non singular in various regions of the configuration space \cite{Barrett} (typically, it becomes singular
in the presence of configurations with right angles between the triangles); 
therefore in those regions the dihedral angles are well defined functions of the areas
and it makes sense to consider the following 4-simplex action,
\equ\label{Ham}
S_{\si}(A_t) = \sum_t A_t\, \Big( \pi - \theta_t(A)\Big).
\nequ

\item On a generic triangulation, the situation is more complicated: there are 
more triangles than edges, thus constraints on the area variables are needed
when many 4-simplices are considered,
\equ\label{AreaRegge}
S_{\rm R}(A_t) = \sum_t A_t\, \eps_t(A) + {\cal C}(A_t).
\nequ
The role of the constraints is to guarantee that a tetrahedron shared between two 4-simplicies has the same
geometry in both sets of area variables. Regretfully, the explicit form of these constraints is still unknown
in the literature (see however \cite{Mike, Makela, RuthRegge, Wainwright}),
and pursuing the study of these constraints is an important problem that deserves future work.

\end{itemize}

A key object of interest for us is the hessian matrix of \Ref{Ham}. To keep our discussion general,
we consider a single simplex in any dimension $n$. 
For a single $n$-simplex it is convenient to introduce a notation based on the vertices of the $n$-simplex, 
labeled by $i=1 \ldots n+1$. Then
$l_e=l_{ij}$ denotes the length of the edge $e$ linking the vertices $i$ and $j$, and
$\theta_{ij}$ denotes the (internal) dihedral angle between the $(n-1)$-simplex $i$ (i.e. the one
made of all vertices but $i$) and $j$, with the convention 
$\theta_{ii}\equiv \pi$. We also pose $V$ to be the $n$-volume of the simplex,
$V_i$ the $(n-1)$-volume of $i$, and $V_{ij}$ the $(n-2)$-volume of the 
hinge between $i$ and $j$.\footnote{The double index notation is very useful to deal with an arbitrary dimension
$n$ of spacetime. On the other hand, care is required in 3d, where confusion might arise because
both $l_{ij}$ and $V_{ij}$ are variables for the edge lengths. $V_{ij}$ is the hinge between the triangles
$i$ and $j$, namely the edge \emph{opposite} to $l_{ij}$. Using the notation $\overline{ij}$ to denote the
edge opposite to $l_{ij}$, we have $V_{ij} = l_{\overline{ij}}$. This notation will be useful in section
\ref{Sec4} below.}
As in 4d, an $n$-simplex has the same number
of edges and hinges, $\f{n(n+1)}2$, thus the jacobian matrix $J^{kl}_{ij}\equiv \f{\p V_{ij}}{\p l_{kl}}$ 
is well defined. In the following, we restrict our analysis to configurations 
where hinges and lengths have a non-vanishing jacobian determinant, 
so we can define the $n$-simplex Regge action
\equ\label{Hamn}
S_{\si}[V_{ij}] = \sum_{ij\in \si} V_{ij}\, \Big( \pi - \theta_{ij}(V)\Big).
\nequ
Using the double index notation, the dihedral angles are given by
\equ\label{dihedral}
\sin \theta_{ij}(l_{e}) = \f{n}{n-1} \f{V(l_{e}) \, V_{ij}(l_{e})}{V_i(l_{e}) \, V_j(l_{e})},
\nequ
and $l_e(V_{ij})$ is well defined being det $J^{kl}_{ij}\neq 0$.

The first derivatives of \Ref{Hamn} give simply the dihedral angles,
\equ\label{deriv}
\f{\p S_{\si}}{\p V_{ij}} \equiv \pi - \theta_{ij}.
\nequ
This crucial property can be easily proved from the $n$-dimensional Schl\"afli identity,
\equ\label{schlafli}
\sum_{ij} V_{ij} \, d\theta_{ij} =0.
\nequ
A direct consequence of \Ref{deriv} is that the hessian is equal to the $({n(n+1)}/2\times {n(n+1)}/2)$ jacobian 
matrix for changing variables from $n-2$-volumes to dihedral angles,
\equ\label{1}
Q_{ijkl} \equiv \f{\p^2 S_{\si}}{\p V_{ij} \p V_{kl}} \equiv - \f{\p \theta_{ij}}{\p V_{kl}}.
\nequ
This change of variables is singular: from \Ref{schlafli} we immediately see that $V_{ij}$ is a 
left and right null vector for \Ref{1}.
This turns out to be the only null vector: below we give the explicit expression
for $Q_{ijkl}$ and show that it has rank $\f{n(n+1)}2-1$ and that the
vanishing of its determinant is related to the closure of a flat $n$-simplex.

As det $J^{kl}_{ij}\neq 0$, we can rewrite \Ref{1} as
\equ\label{q}
Q_{ij kl} = - \sum_{mn} \f{\p \theta_{ij}}{\p l_{mn}} \, \f{\p l_{mn}}{\p V_{kl}}.
\nequ
This expression allows us to evaluate $Q$ in two steps, (i) computing the matrix
$\f{\p \theta_{ij}}{\p l_{mn}}$, and (ii) computing the inverse matrix
of $\f{\p V_{mn}}{\p l_{kl}}$. The latter has a simple form in any dimension (see appendix \ref{appendixA}), being non zero only
when $l_{kl}$ belongs to the hinge $V_{mn}$. In particular
in 4d it has only 30 non-zero entries (see for instance \cite{Makela}),
and its determinant is different from zero provided we exclude
configurations including right angles between triangles, as mentioned above.
Then, its inverse entering \Ref{q} is well defined and it can be computed
algebraically. 

The first term $\f{\p \theta_{ij}}{\p l_{kl}}$ is the most interesting part (and furthermore
its analysis also concerns conventional length Regge calculus). It can be explicitly computed
as shown in the Appendix,
\equ\label{bianca}
\f{\p \theta_{ij}}{\p l_{kl}} = \tan \theta_{ij}\,\f{\p}{\p l_{kl}}
\Big(\ln {V} + \ln V_{ij} - \ln V_i -\ln V_j \Big) =
\f1{n^2} \, \f{l_{kl}}{\sin\theta_{ij}}\, \f{V_k V_l}{{V}^2} \, C_{ijkl},
\nequ
where 
\equ\label{A}
C_{ijkl} \equiv \Big(\cos\theta_{ik}\,\cos\theta_{il}+ \cos\theta_{jk}\,\cos\theta_{jl}\Big)\cos\theta_{ij}
+ \cos\theta_{ik}\,\cos\theta_{jl} + \cos\theta_{jk}\,\cos\theta_{il}.
\nequ
Explicitly,
\equ
C = \left( \begin{array}{ccc}
\sin^2\theta_{12} & -\sin^2\theta_{12} \, \cos\theta_{23} & \ldots \\
-\sin^2\theta_{13} \, \cos\theta_{23} & \sin^2\theta_{13} & \ldots \\
\ldots & \ldots & \ldots
\end{array}\right).
\nequ
Notice that the matrix $C_{ijkl}$ is not symmetric under $ij\leftrightarrow kl$. 
Only the full matrix \Ref{q} is symmetric.
Recalling the formula (no sum over indices)
$\det (a_i M_{ij} b_j) = (\prod_{ij} a_i b_j) \det M_{ij}$, the determinant of \Ref{bianca} is
\equ
\det \f{\p \theta_{ij}}{\p l_{kl}} = 
\left(\prod_{ij, kl} \f1{n^2} \f{l_{kl}}{\sin\theta_{ij}} \f{V_k V_l}{V} \right) \det  C.
\nequ
Here $\det C$ is a polynomial of order $(n+1)^2$ in the cosines, 
and it can be shown by direct computation with an algebraic manipulator that
\equ\label{detC}
\det C = \left(\prod_i {G}^*{}_{ii}\right) \det {G},
\nequ
where ${G}_{ij} = \cos\theta_{ij}$ is the Gram matrix 
with the convention ${G}_{ii}=-1$, and ${G}^*{}_{ij}$ its cofactor matrix.
Analogously one can show that all the minors are never singular and thus the rank is $\f{n(n+1)}2-1$.
This result shows that the non invertibility of $Q$ has its root in the vanishing of the
determinant of the Gram matrix on flat spacetime, namely in the 
closure of a flat $n$-simplex.\footnote{On the other hand,
the determinant of the Gram matrix for a spherical or a hyperbolic $n$-simplex does not vanish.
This means that a perturbative expansion around these two types of background would directly
yield an invertible quadratic order. We will comment on this point in the conclusions.} 

The elegant expression \Ref{bianca} is fundamental to understand the connection between classical Regge calculus
and the free propagator of the quantum theory: being
the quadratic order of the expansion of the Regge action around a fixed background,
the matrix $Q$ is the key input to compute the free graviton propagator in Regge calculus, as we show below.
In the special equilateral configuration, \Ref{bianca} reproduces the results derived in \cite{grav} and it is used 
to study the free graviton propagator from spinfoams quantum gravity.

The explicit expression \Ref{bianca} allows us to address two important questions on the hessian matrix:

\begin{itemize}
\item \emph{What is its scaling?}
Recall that the free propagator scales as the inverse squared distance
between the two points, evaluated with respect to the background geometry. In the 4d spinfoam calculations,
these points are associated with centers of the triangles in the boundary of the 4-simplex.
If the hessian $Q_{ijkl} \equiv Q_{tt'}$ is the crucial dynamical input as argued in \cite{grav}, 
then it should encode the distance between the (centers of the) triangles $t$ and $t'$.
In particular, one might naively expect $Q_{tt'}$ to scale as some power of this distance.\footnote{Notice that
this naive intuition could not be checked in the calculations of \cite{grav}, because there 
only the simple equilateral configuration of the 4-simplex was considered,
thus all the distances are the same and the scaling is trivial.} 
The general expression \Ref{bianca} shows clearly that this naive expectation fails: 
even if $Q$ has a simple geometry dependence, the various components of \Ref{q}
can not be related to distances between points (such as centers of triangles) on the boundary of the 4-simplex. 

Nevertheless we will see below that the matrix is indeed related to the free propagator:
the right scaling only emerges when many 4-simplex contributions are glued together.
In the setting of \cite{grav}, this means that the contribution from the boundary state is crucial,
as we discuss below.

\item \emph{What is the meaning of its non invertibility?}
We know that in the continuum the quadratic term of the Einstein--Hilbert action
(to which the Regge action reduces in the continuum limit) is not invertible due to the invariance
under diffeomorphisms. Can the non invertibility of $Q$ have a similar origin?
The null vector $V_{kl}$ does not appear to be related to a local symmetry such as the diffeomorphisms,
but we will see below that it is indeed the remnant of the continuous diffeomorphisms on the single simplex.

Notice that a natural way of making $Q_{ijkl}$ invertible is to add the rank 1 matrix built out of its null vector $V_{kl}$,
\equ\label{Qinv}
\wtl{Q}_{ijkl}\equiv Q_{ijkl} + \xi \, {{V}^{\f {3(2-n)}n}} \, {V_{ij} V_{kl}},
\nequ
where the factor ${V}^{\f{3(2-n)}n}$ is chosen on dimensional grounds, and $\xi$ is a real number.

\end{itemize}

In the next section we use a 3d model to show explicitly that $Q$ is related to the free graviton propagator 
and that its non invertibility comes from the invariance under diffeomorphism of the original continuum action.
In section \ref{SecConcl} below we come back to the 4d case.

\section{3d lattice model}\label{Sec4}

In this section we describe how the Regge action is related to the lattice graviton
propagator developping the original analysis by Hamber and Williams \cite{Ruth3d}. 
We consider the 3d case where the lengths are the fundamental variables, 
thus we can address the questions at the end of the previous
section without the extra complicacy of the constraints \Ref{AreaRegge} needed in 4d. 
[We expect the results of this section to apply also to 4d length Regge calculus, and so to area Regge calculus if this is proved equivalent to length Regge calculus.]
Using \Ref{bianca}, we have for the tetrahedron
\equ\label{bianca3d}
Q^\tau{}_{ijkl} = -\f{\p \theta_{ij}}{\p V_{kl}} = 
-\f19 \, \f{l_{\overline{kl}}}{\sin\theta_{ij}}\, \f{A_{\overline{k}} \, A_{\overline{l}}}{{V}^2} \, C_{ij\overline{kl}}
\nequ
with $C$ the same matrix as before, and the following notation has been introduced:
$\overline{kl}$ is the edge opposite to $kl$, and $A_{\overline{k}} \, A_{\overline{l}}$
the (unique) pair of triangles different from $A_k \, A_l$. Notice that the
need of the opposite edge in the columns of $C$ comes from the fact that in \Ref{bianca3d} we are deriving with respect to $V_{kl}=l_{\overline{kl}}$. 
The double index notation is required only in this formula. For simplicity in the following calculations,
let us introduce the following single index notation: we use $l_e=l_{ij}=V_{\overline{ij}}$ for the edge $e$ between
$i$ and $j$, and $\theta_e = \theta_{\overline{ij}}$ for the dihedral angle associated with the edge $e$ and thus
hinging the pair of triangles $A_{\overline{i}} \, A_{\overline{j}}$. Finally, we define the tetrahedral matrix
$Q^\tau{}_{ee'} \equiv -\f{\p \theta_e}{\p l_{e'}}\equiv Q^\tau{}_{\overline{ij}\overline{kl}}$. This definition helps
the comparison with the results of \cite{Ruth3d}.

The 3d Schl\"afli identity reads $\sum_e l_e \, d\theta_e =0$,
and the null vector is $l_e$. As we discussed above for any dimension, the matrix \Ref{bianca3d} does not have simple
scaling properties with respect to the background geometry.\footnote{The only elements for which 
we were able to find a simple relation to the distances are the ones for opposite edges: using \Ref{dihedral} we have
\equ
Q_{ij \overline{ij}} = - \f19 \f{ l_{ij} }{\sin\theta_{ij}} \f{A_i A_j}{V^2} \sin^2\theta_{ij}=
- \f1{6V} l_{ij} l_{\overline{ij}} = - \f{1}{d_{(ij),(\overline{ij})} \sin \phi_{(ij),(\overline{ij})}},
\nequ
where $d_{(ij),(\overline{ij})}$ is the distance between the midpoints of the two opposite edges $l_{ij}$
and $l_{\overline{ij}}$, and $\phi_{(ij),(\overline{ij})}$ the angle between them. \label{prova}}
This matrix is the quadratic order of the Regge action on a single 
\\ \\ 
\begin{minipage}{11cm}\vspace{-0.25cm}
tetrahedron. To study how $Q^\tau$ is related to the free graviton propagator,
we consider an infinite rectangular lattice, and divide each block into six tetrahedra as shown in the picture, namely
drawing the diagonals on the six exterior faces plus the interior diagonal
connecting the vertices 0 and 7. All tetrahedra contain the vertices 0 and 7, and can be obtained
from the six possible ways of going from 0 to 7 along three cartesian edges. Consequently each tetrahedron has the same 
edge structure: three cartesian edges, two face diagonals, and 
\end{minipage}\hspace{1cm}
\begin{minipage}{4.5cm}
{\includegraphics[width=4cm]{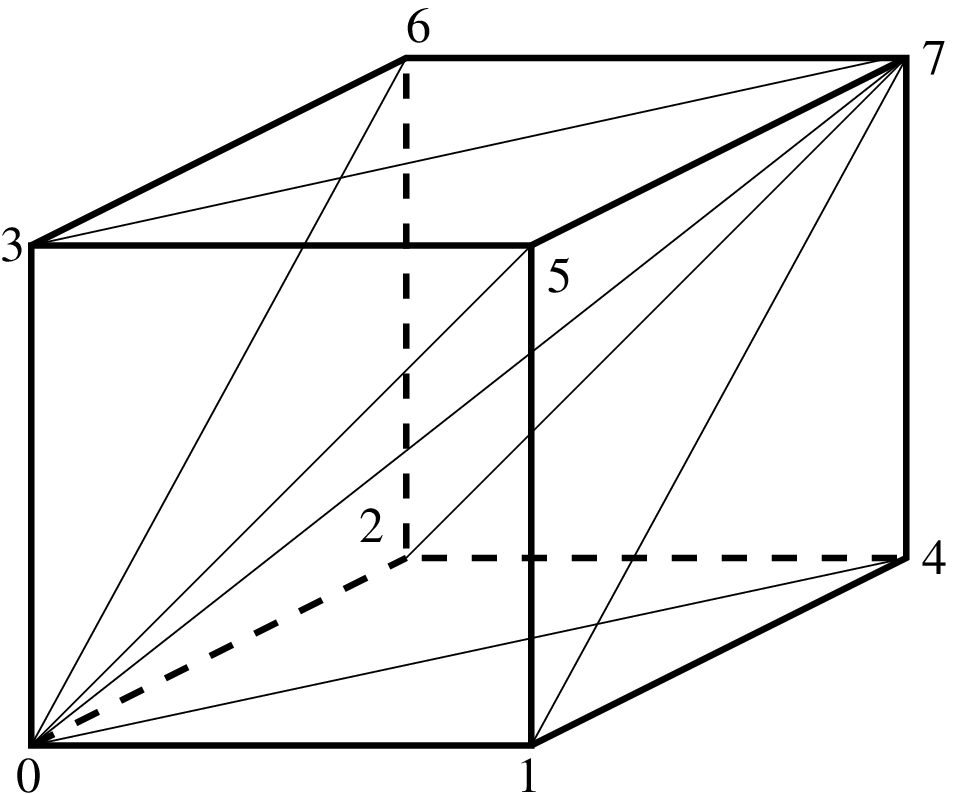}}
\end{minipage}
\\ \\ 
the body diagonal (07).
All the edges are oriented (from smaller to bigger number), so that 
in the full lattice each edge can be identified giving its starting vertex and its direction,
$l_e \equiv l_{\hat e}^v$, where $\hat e=(01) \ldots (07)$ and $v$ are the vertices in the lattice.
Adapting cartesian coordinates to this lattice, we have 
$(01) = \hat x^\mu = (1,0,0)$, $(02) = \hat y^\mu = (0,1,0)$, $(03) = \hat z^\mu = (0,0,1)$, 
$(04) = \hat x^\mu + \hat y^\mu = (1,1,0)$, and so on.

We now introduce the background around which we will compute the correlations.
To keep the lattice regular but also the tetrahedral matrix $Q^\tau$ as general as possible, 
we consider a rectangular background, with lattice spacings 
in the three cartesian directions given respectively by
$l_{{\hat x}}^v = a$, $l_{{\hat y}}^v =b$ and $l_{{\hat z}}^v = c$, for any $v$.
With this background, the six tetrahedra in each block are obtained by permutation of $a,b,c$, theyhave all the same
volume $abc/6$, but the six edges in each tetrahedron are all different and thus
the tetrahedral geometry rather general. Consider for instance the tetrahedron 
$\tilde\tau\equiv0147$: its edge lengths are 
{\small${(a, \sqrt{a^2+b^2}, c, b, \sqrt{b^2+c^2}, \sqrt{a^2+b^2+c^2})}$} and its Gram matrix is
\equ\nonumber
G^{\tilde\tau}_{ij} = \left( \begin{array}{cccc}
-1 & \f{b}{\sqrt{b^2+c^2}} & 0 & 0 \\ 
\f{b}{\sqrt{b^2+c^2}} & -1 & \f{ac}{\sqrt{a^2+b^2}\sqrt{b^2+c^2}} & 0 \\
0 & \f{ac}{\sqrt{a^2+b^2}\sqrt{b^2+c^2}} & -1 & \f{b}{\sqrt{a^2+b^2}} \\ 
  0 & 0 & \f{b}{\sqrt{a^2+b^2}} & -1 
\end{array}\right).
\nequ
where the numbering of row and column is $0,1,4,7$
We use the following notation for the perturbative expansion: $l_e$ for the background values, $\ell_e$ for the variables, and define the fluctuations $\d_e \equiv {\ell_e - l_e},$ for which we also use the notation $\d_{\hat e}^v$ where $\hat e$ denote an oriented edge. We are interested in computing the correlators between fluctuations around this background
to quadratic order. To do so, we evaluate \Ref{bianca3d} on the background, for each of the six tetrahedra in the block.
For instance, the tetrahedron $\tilde\tau\equiv 0147$ gives
\equ\nonumber
l_e l_{e'} {Q}^{\tilde\tau}{}_{ee'} 
={\scriptsize{ \frac{1}{abc}\left( \begin{array}{cccccc}
0 & 0 & 0 & 0 & a^2 c^2 & -a^2 c^2 \\
0 & -b^2 (a^2+b^2) & b^2(a^2+b^2+c^2) & b^2(a^2+b^2) 
& -(a^2+b^2)(b^2+c^2) & a^2 c^2 \\
0 & b^2(a^2 + b^2 +c^2) & -b^2(a^2+b^2+c^2) &  -b^2(a^2+b^2+c^2) 
&  b^2(a^2 + b^2 +c^2)& 0\\
0 & b^2(a^2+b^2) & -b^2 (a^2+b^2+c^2) & -b^4 & b^2 (b^2+c^2)&0\\
a^2 c^2 & -(a^2 +b^2)(b^2+c^2) &  b^2(a^2 + b^2 +c^2) 
& b^2 (b^2+c^2) & -b^2 (b^2+c^2) &0 \\
-a^2 c^2 & a^2 c^2 & 0 & 0 & 0 & 0 
\end{array}\right),}}
\nequ
where the numbering of rows and columns is $(01)(04)(07)(14)(17)(47)$. 
Due to the regularity of the block, the matrices associated to the other tetrahedra can be 
obtained by appropriate permutations.
Then adding up the matrices for the six tetrahedra one obtains the quadratic term of the Regge
action for the block,
\equ\label{defcube}
S_{\rm b}[\d_{e}] = \f12 \sum_{\tau \in {\rm b}}\sum_{e,e'\in \tau} \d_{e} \, {Q}^{\tau}{}_{ee'} \, \d_{e'}
= \f12 \sum_{e,e'\in {\rm b}} \d_{e} \, {Q}^{\rm b}{}_{ee'} \, \d_{e'}.
\nequ
The equation above defines a matrix $Q^{\rm b}$ associated with the block. This matrix is
$(19 \times 19)$, corresponding to the 12 cartesian edges, the 6 face diagonals and the body diagonal.
We report its explicit expression in the Appendix (see \Ref{Qb}).
The total lattice action $S_{\rm L}$ is obtained summing over the blocks of the lattice, and
it can conveniently written as a sum over vertices, 
\equ \label{Qv}
S_{\rm L}[\d_{e}] \equiv 
\sum_{\rm b} S_{\rm b}[\d_{e}] = \f12 \sum_{v, v'} \sum_{\hat e\in v} \sum_{\hat e'\in v'} 
\d_{\hat e}^v \, {Q}^{vv'}{}_{\hat e \hat e'} \, \d_{\hat e'}^{v'},
\nequ
where $\hat e\in v$ denotes one of the seven edges starting in $v$. 
For an infinite lattice, ${Q}^{vv'}$ is an infinite dimensional matrix non vanishing only for first-neighbour vertices. Remarkably, we find for this matrix the following simple scaling,
\equ\label{defM}
{Q}^{vv'}{}_{\hat e \hat e'} \equiv -\f{l_{\hat e} \, l_{\hat e'}}{abc}\, M^{vv'}_{\hat e \hat e'},
\nequ
where the entries of $M$ are adimensional and independent on $a,b,c$; they are
non zero only for first-neighbour vertices, and furthermore there are no correlations between 
edges starting at $v$ and edges starting at $v+\hat x +\hat y + \hat z$. 
The non zero entries of $M^{vv'}$ are structured in 7x7 blocks reported in the Appendix.

This simple scaling looks surprising (specially if compared with the non transparent scaling behaviour
of the single tetrahedron matrix $Q^{\tau}$), and the reader might wonder if it is a peculiarity of the
regular lattice chosen that would be lost on a non regular lattice. However, we show in Appendix \ref{appendixB}
that the action \Ref{Qv} can be obtained discretizing the quadratic term of the expansion
of the Einstein--Hilbert action around the flat metric $g_{\mu\nu}^{\rm bg} =$ diag$(a^2,b^2,c^2)$ (see Appendix \ref{appendixB}). Remarkably, this discretization will produce exactly the scaling
of \Ref{Qv}. This is a peculiarity of 3d GR that we comment on in the Appendix.

In spite of the appealing scaling, at this stage it is still
unclear how this quadratic action will produce the correct scaling of the propagator.
On the other hand, in the equilateral configuration $a=b=c$ this block can be immediately recognized as 
the building block of the action term in \cite{Ruth3d}, which is there shown to lead to the correct free
propagator. In the following, we simplify\footnote{In particular, we 
do not use trace reversed variables as in \cite{Ruth3d}.} 
and generalize the procedure of \cite{Ruth3d} to obtain the free propagator on this rectangular lattice.

As in the continuum, the quadratic action \Ref{Qv} 
has local symmetries that need to be gauge fixed in order to compute the free propagator. 
Calling $S_{\rm gf}$ the gauge fixing term, we have
\equ\label{W3d}
W_{ee'} = \int \prod_i d\d_i \ \d_e \, \d_{e'} \, e^{-S_{\rm L}[\d_e]-S_{\rm gf}[\d_e]}.
\nequ
The standard procedure at this point would be to choose a particular gauge fixing,
and going to Fourier variables to diagonalize the kinetic term.
However, the use of a simplicial lattice leads to the presence of additional spurious variables
in the action on top of the gauge degrees of freedom. This comes
from the fact that the simplicial lattice gives seven directions
for each vertex in the lattice, as opposed to the six variables per point of the metric.\footnote{
We stress that this is an issue of the simplicial lattice alone and not
of the theory being considered here, and indeed it would be present even
if one was studying lattice gauge theory on a simplicial lattice.} 
To isolate the spurious quantities, we make the following change of variables,
\eqa \label{hnew}
h_{\hat e}^v \equiv \sum_{(v', {\hat e'})} T^{vv'}_{{\hat e}{\hat e'}} \, \d_{\hat e'}^{v'},
\neqa
where $T^{vv'}$ is defined in the Appendix (see \Ref{T}) and has entries of length dimension one.
The entries vanish for non first-neighbour vertices, thus this change of variables is local 
in the sense that it mixes only $\d_{\hat e}^v$ which are first neighbours.
The new variables $h_{\hat e}^v$ have length dimension 2, and can be obtained discretizing the continuum metric fluctuations $h_{\mu\nu}(x)$ with $x\mapsto v$ and $(\mu\nu)\mapsto \hat e$ according to
the index correspondence $(11)(22)(33)(12)(13)(23)\mapsto 123456$. This index correspondence
defines a matrix $A^{\mu\nu}_{\hat e}$ such that $h^v_{\hat e} = A^{\mu\nu}_{ \hat e}h_{\mu\nu}(v)$, and
selects $h_7^v$ as the spurious variable. 
In these new variables the action (\ref{Qv}) takes the form 
\equ\label{baction3}
S[h^v_{\hat e}]= \f12 \sum_{{v, v'} \atop {1\leq \hat e,{\hat e'}\leq 6}} 
 h^v_{\hat e} \, L^{vv'}_{\hat e{\hat e'}}\,  h^{v'}_{\hat e'} - \frac{1}{abc}\sum_{v} ( h^v_7)^2,
\nequ
where we see that $ h^v_7$ completely decouples from the others.
We can integrate out this spurious variable 
and ignore it in the following, that is from now on $\hat e=(1,\ldots,6)$.\footnote{This means
that we do not look at correlations including the spurious variables. 
However, notice that such correlations vanish in the continuum limit $(a,b,c)\mapsto 0$
homogeneously.}
The explicit expression of $L^{vv'}$, which is rather lengthy, can be straightforwardly worked out from $Q^{vv'}$ and $T^{vv'}$ given in the Appendix. Remarkably but not surprisingly, it can be also obtained by a discretization of   
the continuum Einstein--Hilbert action $\sqrt{g}R$ around the background metric $g^{\rm bg}=\text{diag}(a^2,b^2,c^2)$, that is 
\equ\nonumber
\f12 \int d^3 x \sqrt{g^{\rm bg}}
\left(
\f12h^\mu_\mu \partial^\nu \partial_\nu h^\sigma_\sigma
-\f12 h_{\mu\nu}\partial_\sigma\partial^\sigma h^{\mu\nu}+
h_{\mu\nu}\partial_\sigma \partial^\nu h^{\mu\sigma }-
h_{\mu\nu}\partial^\mu\partial^\nu h_\sigma^\sigma
\right) \simeq
\f12 \sum_{{v, v'} \atop {1\leq \hat e,{\hat e'}\leq 6}} 
 h^v_{\hat e} \, L^{vv'}_{\hat e{\hat e'}}h^{v'}_{\hat e'}
\nequ
where $h_{\hat e}^{v} = A_{\hat{e}}^{\mu\nu} h_{\mu\nu}(v)$. The lattice derivatives are defined in
the conventional way; denoting $\hat \mu$ the unit vector 
in the direction $\mu$ and $l_\mu=(a,b,c)$ its length, we have
$\Delta_\mu \phi(v) \equiv [\phi(v+\hat \mu)-\phi(v)]/l_\mu \simeq \p_\mu \phi(v)$.

$L^{vv'}$ has six variables and three zero modes per vertex,
that we also report in the Appendix (see \Ref{traslnull}).
These three zero modes correspond to translations in the three cartesian directions, and are the
remnants of the continuous diffeomorphism invariance on the lattice, 
as we discuss in the next section.
This gauge symmetry can be fixed by adding a suitable term to (\ref{baction3}),
replacing the matrix $L^{vv'}_{\hat e{\hat e'}}$ by 
$\wtl L^{vv'}_{\hat e{\hat e'}} = L^{vv'}_{\hat e{\hat e'}}+ L^{\rm gf}{}^{vv'}_{\hat e{\hat e'}}$.
As $L^{vv'}_{\hat e{\hat e'}}$ comes from the discretization of the continuum action,
it is useful to construct also the extra term $L^{\rm gf}{}^{vv'}_{\hat e{\hat e'}}$ from the
discretization of a continuum gauge fixing term. In particular it is convenient to 
choose the harmonic gauge as in \cite{Ruth3d}, 
because it leads to a simple expression for the propagator.\footnote{An
interesting question, which we do not address here, is to consider other gauges such as
the radial gauge suggested in \cite{Magliaro}, and see
how they affect the final expression of the lattice propagator.} 
The harmonic gauge fixing term is $\f12 \sqrt{g^{\rm bg}}  \, C^\mu C_\mu$ with 
$C_\mu \equiv \p^\rho h_{\mu\rho} - \f12 \p_\mu h^\rho{}_\rho$, where the indices are raised and 
lowered with the background metric. We report the discretization of $C_\mu$ in the Appendix
(see \Ref{C}). Using this gauge fixing, the matrix elements of $\wtl L$ are given by
\equ\label{finito}
\wtl L^{vv'}_{\hat e{\hat e'}} = {abc} \sum_{\hat f{\hat f'}} \, 
D_{\hat e{\hat f}} \, \Delta^{vv'}_{\hat f{\hat f'}} \, D_{\hat f'{\hat e'}}
\nequ
where 
\equ
D \equiv {\rm diag}\Big(
a^{-2}, b^{-2}, c^{-2}, (ab)^{-1}, (ac)^{-1}, (bc)^{-1}\Big),
\nequ
and
\equ\label{baction5}
\Delta^{vv'} \equiv \f\Delta4
{\tiny { \left( \begin{array}{cccccc}
1 &  -1 & -1 &0&0&0 \\
 -1 & 1 & -1 &0&0&0 \\
- 1 &  -1 & 1 &0&0&0 \\
0&0&0& 4 &0&0 \\
0&0&0& 0 &4 &0\\
0&0&0& 0&0& 4
\end{array}\right)}}, \qquad 
\Delta \equiv  \sum_{\hat \mu} \f1{l_{\mu}^2} 
\Big( \d_{v',v+\hat \mu}+ \d_{v',v-\hat \mu} - 2 \d_{v',v}\Big). 
\nequ
The quantity $\Delta$ can be immediately recognized as the laplace operator on the rectangular lattice 
here considered.
Using the same $\mu\nu \mapsto \hat e$ correspondence as above,
it is easy to show that the kinetic term \Ref{finito} reduces in the continuum limit to the quadratic term in the harmonic gauge, 
\equ\label{grav}
S_{\rm harm}[h] = \f12 \int d^3x \ \sqrt{g^{\rm bg}} \ h_{\mu\nu}\, P^{\mu\nu\rho\sigma}\, \Delta \, h_{\rho\sigma},
\qquad P_{\mu\nu\rho\sigma} \equiv \f12 \, {g^{\rm bg}_{\mu\rho} \, g^{\rm bg}_{\nu\sigma} 
- \f14 \, g^{\rm bg}_{\mu\nu} \, g^{\rm bg}_{\rho\sigma}}
\nequ
(here $\Delta$ is the continuum laplace operator).

At this point one evaluates the correlations in the new variables,
\equ\label{W3dh}
W^{v_1v_2}_{\hat e_1\hat e_2} = 
\int \prod_{v,\hat e} d h_{\hat e}^v \ h_{\hat e_1}^{v_1} \, h_{\hat e_2}^{v_2} \, 
e^{-S_{\rm L}[h_{\hat e}^{v}]-S_{\rm gf}[h_{\hat e}^{v}]} =
\f12 \f{g^{\rm bg}_{\mu\rho} \, g^{\rm bg}_{\nu\sigma} + g^{\rm bg}_{\mu\sigma} \, g^{\rm bg}_{\nu\rho} 
- 2 \, g^{\rm bg}_{\mu\nu} \, g^{\rm bg}_{\rho\sigma}}{d(v_1,v_2)} \,
A^{\mu\nu}_{ \hat e_1 } \, A^{\rho\sigma}_{ \hat e_2}
\nequ
obtaining the correct scaling with the distance $d(v_1,v_2)$ between the two points and 
the correct tensorial structure of the 3d euclidean free
propagator.\footnote{We remind the reader the 3d GR without matter has no local degrees of freedom.
Consequently this propagator is a pure gauge quantity, and does not represent a physical particle.}
We conclude that we recover the correct lattice propagator, generalizing the results of \cite{Ruth3d}
to the rectangular lattice.
The calculations of this section show how the single tetrahedron matrix \Ref{bianca3d}
adds up correctly to the lattice kinetic term of linearized Regge calculus.

\section{On the gauge fixing}\label{secgf}
This connection with linearized Regge calculus allows us also to
clarify the origin of the non invertibility of \Ref{q}.
Let us go back to the three zero modes per vertex of the full lattice term $L^{vv'}_{\hat e{\hat e'}}$.
These zero modes reported in the Appendix in \Ref{traslnull} are the three
independent translations that can be performed at each vertex of the lattice, as shown 
also in \cite{Ruth3d} for the cubic case. This is the remnant
of the continuous diffeomorphism invariance on a regular lattice, and consistently in the previous
section we gauge fixed this symmetry simply discretizing the 
harmonic gauge for the continuum (linearized) diffeomorphism invariance.

As an intermediate step towards the symmetries of the single tetrahedron matrix, we can look
at a finite lattice built out of eight blocks. Such a lattice has an interior vertex. The
quadratic action for this lattice (built as in \Ref{defcube} adding up the tetrahedral
matrices in the lattice)
has the zero modes corresponding to the possible translations,\footnote{Namely 
three zero modes corresponding to translations of the (edge variables starting at the) interior vertex, 
two zero modes for
the 2d translations of the vertices on the exterior faces of the lattice, one zero mode for the 1d translation of each
vertex on the exterior edges of the lattice.} plus three additional zero modes corresponding to 
independent global rescaling of the lattice in the three cartesian directions. The latter are
\equ\label{rescale}
\d_{{\hat x}} \mapsto \d_{{\hat x}}+N_{\hat x} \, a, \qquad 
\d_{{\hat y}} \mapsto \d_{{\hat y}}+N_{\hat y} \, b, \qquad \d_{{\hat x+\hat y}} \mapsto 
\d_{{\hat x+\hat y}}+ N_{\hat x} \, \f{a^2}{\sqrt{a^2+b^2}} + N_{\hat y} \, \f{b^2}{\sqrt{a^2+b^2}},
\nequ
and so on. Here $N_{\hat x}$, $N_{\hat y}$ and $N_{\hat z}$ are three free parameters. This
symmetry is a consequence of the finiteness of the lattice.

Next, we reduce the lattice to a single block, and look at the symmetries of the $19\times19$
matrix \Ref{defcube}. We find that all the zero modes related to translations are gone
(the edge variables cannot be translated any longer), but
the global rescalings \Ref{rescale} are still present. These are the three zero modes of \Ref{defcube}
reported in the Appendix (see \Ref{nullscaling}).
Finally, the special case $N_{\hat x} = N_{\hat y} = N_{\hat z} \equiv N$ 
gives the only zero mode that survives the initial tetrahedral matrix $Q^\tau$,
corresponding to the transformation $\d_e \mapsto \d_e + N \, l_e$.

The analysis above allows us to make contact between
the symmetry of $Q^\tau$ and diffeomorphisms. It shows that the zero mode of $Q^\tau$
can be gauge fixed discretizing the continuum gauge fixing term for the diffeomorphism invariance.
For instance, one can use the harmonic gauge fixing for the full lattice described 
in the previous section to induce a gauge fixing term for $Q^\tau$. 
Notice that this would produce a rather different term than the gauge fixing 
discussed in section \ref{Sec2} which, adapting \Ref{Qinv} to 3d, reads
\equ\label{3dgf}
Q_{ee'}^{\rm gf} \equiv \xi \f{l_e \, l_{e'}}{V}.
\nequ
Indeed, the process of gauge fixing $Q^\tau$ alone (for instance using \Ref{3dgf})
and then adding up the gauge fixed matrices as in \Ref{Qv} to obtain a gauge fixed total action
is ambiguous. This is because gluing tetrahedra together
produces additional gauge degrees of freedom, of which a simple collection of copies of \Ref{3dgf}
would not take care. This is evident from the presence in the gauge fixing term $C$ 
(see the Appendix, \Ref{C}) of terms connecting edges belonging to different tetrahedra.
From this perspective the use of gauges without derivatives, such as the radial gauge suggested in
\cite{Magliaro}, might be more promising when trying to relate the tetrahedral gauge fixing to the full lattice
one.\footnote{A way to characterize the lattice gauge fixing in terms of the null vectors, as it
is done in \Ref{3dgf}, is the following. If we denote by $W$ the matrix constructed from the three null vectors 
\Ref{traslnull} corresponding to the translations,  
one can show that $\sqrt{g^{\rm bg}} \, W C^\mu C_\mu W = \Delta \, {\mathbbm 1}$. 
This connection between the gauge fixing term and the null vectors could be used to identify
the gauge fixing term in a sub block of the lattice. It suggests that the gauge fixing is dual to the null vectors,
and so to replace \Ref{3dgf} by $\xi \sqrt[3]{V} {l_e^{-1} l_{e'}^{-1}}$. 
However the argument can not really be completed, as the discrete laplace operator appearing 
above is not well defined on a single tetrahedron.}

\section{General boundary formalism}\label{SecGB}
In the previous sections, we used a 3d model to show how the matrix $Q^\tau$ is related
to the free graviton propagator, and how its non invertibility comes from diffeomorphism
invariance. 
To make contact with the spinfoam calculations of the graviton propagator \cite{grav},
we use the general boundary formalism \cite{Oeckl, Conrady}.
As explained in \cite{Conrady} (and here adapted to 3d), 
we introduce a 2d closed surface $\Sigma$ and split the integration of \Ref{W3d}
in the three regions of spacetime defined as the exterior of the surface, the surface itself, and the
interior. In doing so we write the action as
$S_{\rm R} = S_{\Sigma} + S_{\rm out}$
where both pieces have a boundary term as described in \Ref{Regge1}. In particular
we can choose $\Sigma$ to be the surface of a single tetrahedron $\tau$, so that
$S_{\rm R} = S_\tau + S_{\rm out}$. We can then write \Ref{W3d} as
\equ\label{Wtau}
W_{ee'} = \int \prod_{i\in\tau} d \d_i \, \d_e \, \d_{e'} \, e^{-S_{\tau}[\d]} \, \Psi_\tau[\d],
\nequ
where $i\in\tau$ means the edges belonging to $\tau$, and the integration 
over the edges $i\notin\tau$ outside $\tau$ defines the boundary state
\equ\label{psitau}
\Psi_\tau[\d] \equiv \int \prod_{i \notin \tau} d \d_i \, e^{-S_{\rm out}[\d]}.
\nequ
At this point we can proceed evaluating \Ref{psitau} first, then plugging it into \Ref{Wtau} and 
evaluating the remaining integrals. As all we did is a splitting of the initial
integration \Ref{W3d}, we expect to obtain the same result of section \ref{Sec4}. This formulation is 
nonetheless interesting because it seems more suitable for background independent approaches.
In particular in \cite{grav} it is argued that a quantity like \Ref{Wtau} -- and its 4d equivalent 
for area variables, see below -- emerges as the leading
order of the spinfoam graviton propagator.\footnote{In particular, it emerges 
from the simplest spinfoam, the one characterized by a single vertex.
More general spinfoams can be put in correspondence with different choices of $\Sigma$.}
As \Ref{Wtau} gives the right propagator, this emergence would support the correctness
of the semiclassical limit of spinfoams. However in \cite{grav} the boundary state $\Psi_\tau$ is not evaluated
using the definition \Ref{psitau}, but an ansatz for it is made.
To shed light on $\Psi_\tau$ and improve the understanding of the ansatz and of the spinfoam calculations
in general, we focus on \Ref{psitau} in the rest of this section.

The boundary state is a function of the fluctuations $\d_e$ on the boundary of $\tau$. For the linearized theory,
it is a gaussian in the fluctuations, and it is unique once the background and the gauge are fixed.
It can be evaluated following the same procedure of the previous sections, with a few caveats that 
we stress below. The first difference concerns the expansion of $S_{\rm out}$ around the background.
In expanding \Ref{W3d}, we used the fact that the linear term $\f{\p S_{\rm R}}{\p l_e} \equiv \eps_e$ 
vanishes on the background solution, so that the leading order is quadratic. 
Yet this time, because both $S_\tau$ and $S_{\rm out}$
have a boundary contribution, the linear term does not vanish: it gives the extrinsic geometry of the boundary,
according to \Ref{deriv}. For instance for $S_\tau$ we have
\equ\label{pluto}
S_\tau[\d_e] = S^{(0)}_\tau - \sum_{e\in \tau}\theta_e \, \d_e + \f12\, \sum_{e\in\tau}Q^\tau_{ee'} \, \d_e \, \d_{e'},
\nequ
where the dihedral angles $\theta_e$ of $\tau$ give its extrinsic geometry, and 
the quadratic term is the matrix \Ref{q}. The expansion is analogous for $S_{\rm out}$, in particular
the linear term will be exactly the same (up to orientation), being $\tau$ also the boundary of the outside region.
At first sight the presence of linear terms might look like a big difference with the calculation of the previous section, but this difference is artificially created by the splitting of the original integration
of \Ref{W3d} in two parts as in \Ref{Wtau}. As (the background geometry of) $\tau$ is adapted within a flat triangulation, the dihedral angles of these linear terms in \Ref{psitau} add up to give zero deficit angles
after the integration over the lengths, and finally restore the structure of \Ref{W3d}.
This shows that it is natural to expect a linear term in the boundary state.

A second caveat in the evaluation of the boundary state concerns how to deal with the spurious variables.
In section 3 we isolated them performing the transformation \Ref{hnew}. This transformation links
only first-neighbours in the rectangular lattice, thus it is local on a box, but it is not well-defined on a single
tetrahedron. This is not a problem as a tetrahedron has only six variables and so no
spurious ones, but for a generic $\Sigma$ one has to find a procedure to deal with this issue.
In the special case when $\Sigma$ is an infinite plane, the spurious variables can be isolated 
again by a change of variables such as \Ref{hnew}. This case was considered in \cite{Hamber},
where it was shown that the quadratic order of the boundary state for 4d length Regge calculus
reduces in the continuum limit to the correct expression (given for instance in \cite{Kuchar, Mattei}). 

The last caveat concerns the gauge fixing. Evaluating the integrals defining \Ref{psitau} requires a gauge fixing, thus the boundary state is gauge dependent. This point is particularly 
relevant if one is given the explicit boundary state (for instance through an ansatz)
and plugs it into \Ref{Wtau} to compute the propagator: to get the correct answer, 
the remaining integrations in \Ref{Wtau} have to be gauge fixed consistently with the choice of gauge for the 
boundary state.
Notice that contrarily to the continuum case, the gauge chosen
might not be transparent in the lattice explicit boundary state. The most natural procedure to avoid this problem
is to split the full lattice gauge fixing term into boundary and bulk contributions.

\section{Conclusions and perspectives on the spinfoam \\ calculations of the graviton}\label{SecConcl}
In this paper, we studied the connection between the hessian matrix of the Regge action for an  
$n$-simplex and the free propagator of the linearized quantum theory.
Our first result is the general formula for the hessian matrix, given in \Ref{bianca}. 
This is needed to compute the perturbative expansion of the Regge action around an arbitrary background,
and enters the evaluation of the free propagator, or 2-point function. 
The latter scales as the background distance between 
the two points to the power $2-n$, so one might naively expect a similar scaling for the hessian matrix.
Contrarily to these expectations, the hessian shows no transparent scaling
(with the only exception described in footnote \ref{prova}). 
Furthermore, we showed that the hessian has a single zero mode, and thus is not invertible.
We computed explicitly the determinant of the hessian, equation \Ref{detC}, and showed that the
zero mode has its root in the flatness of the $n$-simplex.

To understand both the scaling and the non invertibility of the hessian, we considered a 3d 
lattice model.
We constructed the full lattice action adapting \Ref{bianca} to a rectangular lattice.
Then we generalized (and simplified) the techniques introduced in \cite{Rocek} and \cite{Ruth3d}
and show that upon properly gauge fixing the action one obtains the correct free lattice graviton propagator,
given in \Ref{W3dh}. The procedure answers the issues raised above in the following way.
First, the correct distance scaling, hidden in the single tetrahedron term $Q^\tau$, emerges when
the full lattice is considered (from the point of view of the general boundary proposal \Ref{Wtau}, this means only
when the correct boundary state is added).
Second, the zero mode of the hessian is explicitly related to the diffeomorphism invariance of the continuum
action. In particular, we showed that the hessian matrix can be gauge fixed 
discretizing the continuum gauge fixing term for the diffeomorphisms. A consistent gauge fixing is
crucial to obtain the correct free graviton propagator.

These results were obtained explicitly in 3d, but can be extended to 4d length Regge calculus.
On a hypercubical lattice, it was shown in \cite{Rocek} that linearized quantum Regge calculus gives
the correct free spin-2 propagator in the continuum limit. 
To relate those results to the hessian, one can use the 4d $Q^\si$ to construct the full quadratic action
for a 4d rectangular lattice, where the 4-simplices will have all different edge lengths.
Then one can proceed as in section \ref{Sec4} to generalize the results of \cite{Rocek}
to this rectangular lattice.
The calculations become lengthier, but there are no technical obstructions. 

The situation is different if one uses the areas as fundamental variables, as suggested by the
spinfoam calculations. Studying linearized area Regge calculus (and its propagator)
is a more complicate problem, due to the lack of knowledge about the constraints needed
to treat the areas as independent variables, as discussed in section \ref{Sec2}. 
An intermediate step considered in \cite{RuthRegge, Wainwright} is to
take the areas as functions of the edges, $A_t(l_e)$, namely
one assumes to have solved the constraints. Accordingly one looks at dependent fluctuations 
$\d_t(l_e)$. The resulting theory can be shown to be equivalent to conventional length Regge calculus.
In order to extend our conclusions to area Regge calculus, it is crucial to study the constraints
in \Ref{AreaRegge}. Due to the lack of knowledge about their explicit form,
we are not in a position to draw for area Regge calculus the same conclusions we have for the
conventional theory.

Nevertheless the results here obtained help clarify important aspects of the spinfoam graviton 
calculations of \cite{grav}, in particular the issue of gauge fixing.
Let us summarize the steps from \Ref{Wtau} to the final propagator expression \Ref{W3dh}:
(i) we expand $S_\tau$ to second order around the background;
(ii) we gauge fix its symmetry; 
(iii) we evaluate the boundary state \Ref{psitau}, again at
quadratic order, in the same gauge; (iv) the remaining integrations in \Ref{Wtau} 
give the correct free propagator.

If we compare these steps with the general setting of the calculations in \cite{grav}, the step (i)
is the same, but the steps (ii) and (iii) are not performed.
Concerning (ii), there is no gauge fixing of the symmetry of $S_\tau$. 
Concerning (iii), the boundary state is not evaluated, but rather a gaussian ansatz is made.
This gaussian ansatz has a linear and a quadratic term. The linear term is perfectly
consistent with the one found in the previous section, although in \cite{grav}
its presence is justified on different grounds.\footnote{Notice that
in \cite{grav} the linear term is imaginary, thus giving a phase to
the boundary state. The factor $i$ that makes the linear term into a phase can be understood 
recalling that in the spinfoam calculations one is using a kernel $\exp\{-i S\}$ as opposed to $\exp\{ -S\}$ used here.} The quadratic term, on the other hand, is left generic,
with free parameters giving the spread of the gaussian.
As a consequence, those calculations can not be used to check the correct distance dependence
of the propagator, nor its tensorial structure, because both aspects rely heavily on the precise structure
of the boundary state. Based upon a generic boundary state, the calculations can only test the correct 
overall scaling of the propagator, and this is the result indeed achieved in \cite{grav}.
In principle, one can impose the correct result to constrain the free parameters entering the
gaussian. However this is rather subtle for the presence of the gauge fixing term in the boundary state.
In fact, concerning the absence of gauge fixing in \cite{grav}, point (ii), 
the reader might wonder why those calculations achieve a finite
answer. The reason is simply that the genericity of the boundary state acts as a gauge fixing term
for $Q^\tau$, making the overall (boundary plus bulk) quadratic term of the action invertible. 
Therefore the free parameters appearing in \cite{grav} should be fixed not only by a dynamical
derivation of the ansatz, but also by making a precise choice of gauge for the full lattice.
This is an unusual gauge fixing procedure: in order to get the right
answer, the boundary state must carry also the correct additional gauge fixing information for the bulk term.
A more natural procedure that emerges
from the results presented here is to first choose a gauge for the boundary state, then evaluate it
through its definition \Ref{psitau} (or alternative ways\footnote{It is useful to
remind the reader that, at least in the conventional context of background dependent
quantum field theory, the boundary state can be directly extracted from the kernel.
This procedure for linearized gravity is described in \cite{Mattei}.}), 
and finally fix the symmetry of the bulk action in the same gauge. 

Since the gauge fixing is hidden in the boundary state it seems
important to us to explore the dependence of the boundary state on the gauge
fixing in a more systematic manner at least in Regge calculus. The setup
which we presented in this work appears to be a good starting point for this
task. 
Possible extensions which are relevant for the spinfoam calculations are the use
of different gauge fixings and even less regular lattices.

Understanding these issues for the graviton is particularly relevant since we are computing a gauge
dependent observable. Another strategy to study the semiclassical limit of spinfoams could be to 
look at gauge invariant observables and compare them with their counterpart in
Regge calculus. For instance it would be interesting to adapt to this discrete Regge setting the framework
to study GR perturbation theory in a gauge invariant way developed in \cite{Dittrich}.

A delicate issue to be understood in the general context concerns the spurious variables.
As we have seen identifying them is crucial to reproduce correctly the propagator.
The spurious variables come from the simplicial nature of the lattice, and have nothing to do
with gauge variables, thus fixing the gauge does not take care of them
and they have to be identified in an independent way.
In the regular lattice considered here we were able to do so studying the special way $\d_7^v$
enters the action, but on a general irregular lattice such as the one arising in a general spinfoam 
model this identification could be a very challenging issue.

Another interesting point that could be further developed is the perturbative expansion
around a constant-curvature background. As it is clear from \Ref{detC}, the hessian becomes invertible
if the background is a spherical or hyperbolical $n$-simplex (respectively positive or negative 
Gram determinant). We leave this point open, and make the simple following remark.
As we are dealing only with riemannian signatures, these backgrounds can be associated with
the perturbative expansion in the presence of a non-zero cosmological constant $\Lambda$. 
In fact, $\Lambda>0$ (respectively $\Lambda<0$) can be associated with a global spherical (hyperbolical)
topology. Then this spacetime could be realized on a lattice made out of spherical (hyperbolical)
$n$-simplices. In this case, the $n$-simplex hessian is invertible, 
and gauge degrees of freedom in the full lattice action only arise from the gluing of the
$n$-simplices.

\section*{Acknowledgments}
We thank Carlo Rovelli and Ruth M. Williams for discussions and suggestions.
Research at Perimeter Institute for Theoretical Physics is supported in
part by the Government of Canada through NSERC and by the Province of
Ontario through MRI.

\appendix
\section{Explicit evaluation of $Q$} \label{appendixA}
In this Appendix we report the calculations leading to \Ref{bianca}.
It is convenient to use the affine (or barycentric) coordinates introduced in \cite{Sorkin} for Regge calculus 
(see also \cite{bianca1}). These are defined as follows.
Let $\vec{v}_j,\, j=1,\ldots,n+1$ be vectors from some arbitrarily chosen point in $\mathbb{R}^n$ to the $n+1$ vertices of an $n$-simplex $\sigma$.
Then
\eqa
{\mathbf e}_j \equiv \vec{v}_j- \frac{1}{n+1}\sum_{k=1}^d \vec{v}_k
\neqa
define the affine basis. Notice that $\sum_j{\mathbf e}_j=0$, so this basis is overcomplete. 
As a consequence, the affine coordinates $\tilde x^j$ of a vector $\vec{x}=\sum_j \tilde x^j {\mathbf e}_j$ 
are not unique. Uniqueness can be obtained requiring $\sum_{j}\tilde x^j=0$.

We can also introduce a dual affine basis ${\mathbf e}^j$ defined by
\eqa
\mathbf e_j \cdot \mathbf e^k = \tilde \d^k_j \equiv \d^k_j-\frac{1}{n+1}
\neqa
where $\d^k_j$ is the Kronecker delta. The dual basis satisfies $\sum_j {\mathbf e}^j=0$. 
This enables us to define affine components of an arbitrary tensor in $\mathbb{R}^n$ by contracting it with the appropriate product of affine basis vectors ${\mathbf e}_j, {\mathbf e}^k$.

The key quantity for our calculations is the metric tensor in these coordinates. It can be shown that
the affine components of the metric tensor are \cite{Sorkin}
\equ\label{affinemetric}
\tilde g_{ij}=-\f12 \sum_{k, l} l^2_{kl} \, \tilde \d^k_i \, \tilde \d^l_j.
\nequ
In particular one can easily check that contracting \Ref{affinemetric} with
the edge vectors $(\vec{e}_{ij})^k \equiv \d^k_i-\d^k_j$ gives indeed the length $l^2_{ij}$ of these vectors. 
Laplace's formula for the determinant of $\tilde g_{ij}$ gives the squared of the $n$-volume of the simplex,
\equ\label{volume1}
V^2=\frac{1}{(n!)^{3}}\, \tilde \epsilon^{k_1 k_2\cdots k_n}\,\tilde\epsilon^{l_1 l_2 \cdots l_n}\,\tilde g_{k_1 l_1}\cdots \tilde g_{k_n l_n},
\nequ
where the affine epsilon tensor is defined as
\eqa
\tilde \epsilon^{j_1\cdots j_n} =
\begin{cases}
+1\,\,& \mbox{if the permutation}\,\, j_1 \cdots j_n j_{n+1} \,\, \mbox{is even} \\
-1\,\,  &\mbox{if the permutation}\,\, j_1 \cdots j_n j_{n+1} \,\, \mbox{is odd}
\end{cases}
\neqa
and is vanishing if $\{j_1 \cdots j_n\}$ can not be completed to a permutation of $\{1,\ldots,n+1\}$.

The inverse affine metric components $\tilde g^{ij}$ are defined by 
$\tilde g_{kl} \, \tilde g^{li}=\tilde g^{il} \, \tilde g_{lk}=\tilde \d^i_k$, and one can use \Ref{volume1}
to write
\equ\label{inverse metric}
\tilde g^{ij}
\,=\,\f{n}{(n!)^3}\,\f1{V^2}\,\tilde\eps^{i k_2\cdots k_n}\,\tilde\eps^{j l_2 \cdots l_n}\,\tilde g_{k_2 l_2}\cdots \tilde g_{k_n l_n} \,=\,-\frac{1}{V^2} \frac{\partial V^2}{\partial l^2_{ij}}.
\nequ
The diagonal components of the inverse metric $\tilde g^{ii}$ are 
proportional to the $(n-1)$-volumes of the $(i)$-subsimplices,
\equ\label{invdiag}
\tilde g^{ii}=\f1{n^2}\f{V_i^2}{V^2}.
\nequ
Analogously, the $(n-2)$-volumes $V_{ij}$ of the $(ij)$--subsimplices (i.e. in $4$ dimensions the areas of the triangles not containing the vertices $i$ and $j$) can be obtained by fixing two indices in both epsilon tensors 
of the RHS of \Ref{inverse metric},
\equ
V_{ij}^2 = \f1{((n-2)!)^3} \tilde \epsilon^{ij k_3 \cdots k_n}\tilde \epsilon^{ij l_3 \cdots l_n} \, \tilde g_{k_3 l_3} \cdots \tilde g_{k_n l_n}.
\nequ
From this formula we can immediately write the derivatives with respect to the edge lengths,
\equ
\f{\p V_{ij}^2}{\p l^2_{hm}} = -\f{(n-2)}{((n-2)!)^3} \tilde \epsilon^{ijh k_4 \cdots k_n}\tilde \epsilon^{ijm l_4 \cdots l_n} \, \tilde g_{k_4 l_4} \cdots \tilde g_{k_n l_n}.
\nequ

The inverse affine metric can be used to obtain the matrix \Ref{bianca} of derivatives of the dihedral angles
with respect to the edge lengths. To do so, we need to relate the dihedral angles to the inverse affine metric.
The dihedral angle $\theta_{ij}$ is defined to be the angle between the (inner) normals $n(i)$, $n(j)$ to the faces $i$ and $j$. The affine components of these are given simply by
$n(i)_k = \tilde \d^i_k$, and their norm is
\equ
|n(i)|^2 \equiv \sum_l n(i)_k \, n(i)_l\, \tilde g^{kl} = \sum_{kl} \tilde \d^i_k \, \tilde \d^i_l \, \tilde g^{kl} = \tilde g^{ii}
= \f1{n^2}\f{V_i^2}{V^2},
\nequ
where in the last step we used \Ref{invdiag}.
This gives for the dihedral angles
\equ\label{dihedral2}   
\cos\theta_{ij}=-\f{n(i)_k \, n(j)^k}{|n(i)| \, |n(j)|} = 
-\f{\tilde g^{ij}}{\sqrt{\tilde g^{ii} \, \tilde g^{jj}}}.
\nequ
Notice the minus sign, due to the fact that the $\theta$ are defined as the inner dihedral angles. 
This is consistent with the convention $G_{ii}=-1$ for the Gram matrix used in section \ref{Sec2}. 
Using \Ref{invdiag} and \Ref{dihedral2}, we can write the affine inverse metric as
\equ\label{dihedral4}
\tilde g^{ij}=- \f1{n^2} \, \f{V(i)V(j)}{V^2} \, \cos\theta_{ij}.
\nequ

It is now straightforward to obtain the derivative of the dihedral angles with respect to the length variables. First,
we take the derivative of equation (\ref{dihedral2}), getting
\equ\label{dihedral5}
\frac{\partial \theta_{kl}}{\partial l_{hm}}=\frac{l_{hm}}{\sin\theta_{kl}}\frac{1}{\sqrt{\tilde g^{kk}\tilde g^{ll}}}\left(\tilde g^{kh}\tilde g^{ml}+\tilde g^{km}\tilde g^{hl}-\frac{\tilde g^{kl}}{\tilde g^{kk}}\tilde g^{kh}\tilde g^{km}-\frac{\tilde g^{kl}}{\tilde g^{ll}}\tilde g^{lh}\tilde g^{lm}\right).
\nequ
Next, we use (\ref{dihedral4}) to replace the inverse affine metric components by functions of the dihedral angles,
obtaining the final expression in \Ref{bianca}:
\eqa\label{dihedral6}
\f{\p \theta_{kl}}{\p l_{hm}} &=& \f 1{n^2} \, \f{l_{hm}}{\sin\theta_{kl}}\, \f{V_h \, V_m}{V^2}
\Big(\cos\theta_{kh}\cos\theta_{ml}+ \cos\theta_{km}\cos\theta_{hl}+ 
\nonumber \\ && \quad\quad  \quad\quad  \quad\quad  \quad\quad 
\cos\theta_{kl}(\cos\theta_{kh}\cos\theta_{km}+\cos\theta_{lh}\cos\theta_{lm})\Big).
\neqa
Note that this formula holds for general dimensions $n$.

\bigskip

We also describe an alternative derivation of the formula, which gives the intermediate step in \Ref{bianca},
and allows a more geometrical understanding of the different contributions to the differential of the dihedral angles. We start from the general law of sines in any dimension (a derivation can be found at the end of this section),
\equ\label{sine1}
\sin\theta_{ij}=\frac{n}{(n-1)}\frac{V_{ij}V}{V_i V_j }.
\nequ
Taking the total derivative of equation (\ref{sine1}) we obtain
\eqa\label{sine2}
\bd \theta_{ij} &=& \f{n}{(n-1)} \, \f{1}{\cos\theta_{ij}} \, \f{V \, V_{ij}}{V_i \, V_j}
\left( \f{\bd V}{V}+\f{\bd V_{ij}}{V{ij}}-\f{\bd V_{i}}{V_{i}}-\f{\bd V_{j}}{V_{j}}\right) = \nonumber \\
&=& \tan\theta_{ij}
\left( \f{\bd V}{V}+\f{\bd V_{ij}}{V_{ij}}-\f{\bd V_{i}}{V_{i}}-\f{\bd V_{j}}{V_{j}}\right) = \nonumber \\
&=& \f{1}{2} \tan\theta_{ij}
\left(\tilde\delta^k_m+\widetilde P(ij)^k_m -\widetilde P(i)^k_m -\widetilde P(j)^k_m
\right) \tilde g^{ml} \, \bd \tilde g_{kl}.
\neqa
In the last line we used that the volume differentials of the simplex and the subsimplices can be written as
\eqa
\frac{\bd V_x}{V_x}=\frac{1}{2}\widetilde P(x)^{k}_m\, \tilde g^{ml}\, \bd \tilde g_{kl}
\neqa
where $x$ can stand for $\sigma$ (then $V_\sigma=V$) or $i$ or $ij$. 
$\widetilde P(x)^k_m$ for $x=i$ and $ij$ is the projector in affine coordinates onto the subsimplices $(i)$ and $(ij)$, respectively. For $x=\sigma$ we have $\widetilde P(\sigma)^k_m=\tilde \delta^k_m$. These projectors can be expressed in terms of the inverse affine metric components, which leads again to equation (\ref{dihedral5}). From the last line of (\ref{sine2}) one sees easily that the total derivative of the dihedral angle vanishes if contracted with a variation vector $\delta$ corresponding to a global scaling of the simplex, i.e. such that $\bd \tilde g_{kl}(\delta)\varpropto \tilde g_{kl}$.

\bigskip

The law of sine (\ref{sine1}) can be derived by the following consideration \cite{kokkendorff}. Let $d(i)$ be the distance from the vertex $i$ to the subsimplex $(i)$ and $d(i|j)$ be the distance in the subsimplex $(j)$ of the vertex $i$ to the subsimplex $(ij)$. Then we have a right triangle where two of the sides are $d(i)$ and $d(i|j)$ and the angle opposite to the side of length $d(i)$ is the dihedral angle $\theta_{ij}$. Hence we have
\equ
\sin\theta_{ij}=\frac{d(i)}{d(i|j)}=\frac{n}{(n-1)}\frac{V\, V_{ij}}{V_i V_j} 
\nequ
where in the second step we used that for any $m$--simplex $\sigma^m$ we have 
\equ
d(i|\sigma^m)=\frac{m!\,\, V(\sigma^m)}{(m-1)!\,\, V(i|\sigma^m)} \quad .
\nequ
Here $V(\sigma^m)$ is the volume of the simplex $\sigma^m$ and $V(i|\sigma^m)$ is the volume of the subsimplex of $\sigma^m$, that one obtains by removing the vertex $i$ from $\sigma^m$.

\section{3d lattice model}\label{appendixB}
In this Appendix we report explicitly the matrices used in section \ref{Sec4} and give more details on the
calculations. We begin reporting the $19\times19$ block matrix defined in \Ref{defcube}. Adding up all the tetrahedral contributions, we find
\equ\label{Qb}
Q^{\rm b}{}_{ee'}=-\frac{l_{e}l_{e'}}{abc}M^{\rm b}{}_{ee'}
\nequ
where
\eqa\nonumber
{M}^{\rm b}{}_{ee'} 
={\tiny{ \left( \begin{array}{ccccccccccccccccccc}
0&0&0&0&0&0&0& 0&0&-1&0&0&0& 0&0&0&1&1&0 \\
0&0&0&0&0&0&0& 0&0&0&0&0&-1& 0&0&0&1&0&1 \\
0&0&0&0&0&0&0& 0&0&0&0&0&0& 0&0&-1&0&1&1\\
0&0&0&1&0&0&-1& -1&0&1&-1&0&1& 0&0&0&-1&0&0\\
0&0&0&0&1&0&-1& 0&-1&1&0&0&0& -1&0&1&0&-1&0\\
0&0&0&0&0&1&-1& 0&0&0&0&-1&1& 0&-1&1&0&0&-1\\
0&0&0&-1&-1&-1&2& 1&1&-1&1&1&-1& 1&1&-1&0&0&0\\
0&0&0&-1&0&0&1& 1&0&-1&0&0&0& 0&0&0&0&0&0\\
0&0&0&0&-1&0&1& 0&1&-1&0&0&0& 0&0&0&0&0&0\\
-1&0&0&1&1&0&-1& -1&-1&1&0&0&0& 0&0&0&0&0&0\\
0&0&0&-1&0&0&1& 0&0&0&1&0&-1& 0&0&0&0&0&0\\
0&0&0&0&0&-1&1& 0&0&0&0&1&-1& 0&0&0&0&0&0\\
0&-1&0&1&0&1&-1& 0&0&0&-1&-1&1& 0&0&0&0&0&0\\
0&0&0&0&-1&0&1& 0&0&0&0&0&0& 1&0&-1&0&0&0\\
0&0&0&0&0&-1&1& 0&0&0&0&0&0& 0&1&-1&0&0&0\\
0&0&-1&0&1&1&-1& 0&0&0&0&0&0& -1&-1&1&0&0&0\\
1&1&0&-1&0&0&0& 0&0&0&0&0&0& 0&0&0&0&0&0\\
1&0&1&0&-1&0&0& 0&0&0&0&0&0& 0&0&0&0&0&0\\
0&1&1&0&0&-1&0& 0&0&0&0&0&0& 0&0&0&0&0&0
\end{array}\right)}}.
\neqa
The ordering of rows and columns is $(01,02,\ldots,07,14,15,17,24,26,27,35,36,37,47,57,67)$, 
where $ij$ denotes the edge between the vertices $i,j \in (0,1,\ldots,7)$ of the block. 
We find that $Q^{\rm b}{}_{ee'}$ has three null vectors $v_{e}^i,\,\, i=\hat x, \hat y, \hat z $, given by 
\eqa\label{nullscaling}
l_{\hat e} \, v_{\hat e}^{\hat x} &=& (1,0,0,1,1,0,1,0,0,0,1,0,1,1,0,1,0,0,1),  \nonumber\\
l_{\hat e} \, v_{\hat e}^{\hat y} &=& (0,1,0,1,0,1,1,1,0,1,0,0,0,0,1,1,0,1,0),   \nonumber\\
l_{\hat e} \, v_{\hat e}^{\hat z} &=& (0,0,1,0,1,1,1,0,1,1,0,1,1,0,0,0,1,0,0).
\neqa
These null vectors where discussed in section \ref{secgf}, and correspond to the scaling symmetry \Ref{rescale}.

The next step in (\ref{Qv}) is to rewrite the total action as a sum over vertices. For each pair of vertices $(v,v')$ we get a $7\times7$ matrix, with the same scaling as (\ref{Qb}),
encoding the correlations between the seven edges starting from the vertex $v$ and the seven edges starting from the vertex $v'$. We write this as in (\ref{defM}),
\eqa\label{QvApp}
{Q^{vv'}}_{\hat e \hat e'} \equiv -\frac{ l_{\hat e} l_{\hat e'}}{abc} M^{vv'}_{\hat e \hat e'},
\neqa
and report here the relevant elements of $M^{vv'}_{\hat e \hat e'}$. Denoting $\d_{v',v}$ the Kronecker delta
for the lattice vertices, we have
{\footnotesize{
\eqa 
&& \hspace{-1.2cm}
M^{vv'}_{11} = 2 \, \d_{v',v} ,  \quad  M^{vv'}_{12} = \d_{v',v+\hat x+\hat z}+\d_{v',v-\hat y-\hat z} ,  
 \quad  M^{vv'}_{13} = \d_{v',v+\hat x+\hat y}+\d_{v',v-\hat y -\hat z } , \quad 
 M^{vv'}_{14} = - \d_{v',v} - \d_{v',v-\hat y} ,   
 \no  \quad  \quad \no 
&& \hspace{-1.2cm} 
M^{vv'}_{15} = - \d_{v',v} - \d_{v',v-\hat z} ,  \quad M^{vv'}_{16} = - \d_{v',v+\hat x} - \d_{v',v-\hat y-\hat z} , \quad  M^{vv'}_{17} = \d_{v',v-\hat y} + \d_{v',v-\hat z} ,  
 \quad \quad \no  \quad  \quad \no
&& \hspace{-1.2cm} 
M^{vv'}_{21} = \d_{v',v-\hat x-\hat z}+\d_{v',v+\hat y+\hat z} ,  \quad  M^{vv'}_{22} = 2 \, \d_{v',v} ,  \quad
 M^{vv'}_{23} = \d_{v',v+\hat x+ \hat y}+\d_{v',v-\hat x- \hat z} , \quad
 M^{vv'}_{24} = - \d_{v',v} - \d_{v',v-\hat x} , 
 \no  \quad  \quad \no 
&& \hspace{-1.2cm} 
M^{vv'}_{25} = - \d_{v',v+\hat y} - \d_{v',v-\hat x-\hat z} ,  \quad 
 M^{vv'}_{26} = - \d_{v',v} - \d_{v',v-\hat z} ,  \quad M^{vv'}_{27} = \d_{v',v-\hat x} + \d_{v',v-\hat z} ,  
 \quad \quad \no  \quad  \quad \no
&& \hspace{-1.2cm} 
M^{vv'}_{41} = -\d_{v',v} - \d_{v',v+\hat y} ,  \quad  M^{vv'}_{42} = -\d_{v',v}-\d_{v,v'+\hat x} ,  \quad
 M^{vv'}_{43} = -\d_{v',v-\hat z} - \d_{v',v+\hat x+ \hat y} , \quad M^{vv'}_{44} = 2 \, \d_{v',v} ,  
 \no  \quad  \quad \no \quad  
&& \hspace{-1.2cm} 
M^{vv'}_{45} = \d_{v',v-\hat z} + \d_{v',v+\hat y} ,  \quad 
 M^{vv'}_{46} = \d_{v',v-\hat z} + \d_{v',v+\hat x} , \quad
 M^{vv'}_{47} = - \d_{v',v} - \d_{v',v-\hat z} ,  \quad \quad \no  \quad  \quad \no
&& \hspace{-1.2cm} M^{vv'}_{71} = \d_{v',v+ \hat y} + \d_{v',v+\hat z} ,  \quad  M^{vv'}_{72} = \d_{v',v+ \hat x} + \d_{v',v+\hat z} ,  \quad M^{vv'}_{73} = \d_{v',v+ \hat x} + \d_{v',v+\hat y} , \quad
 M^{vv'}_{74} = - \d_{v',v} - \d_{v',v+\hat z} ,   \no  \quad  \quad \no
&& \hspace{-1.2cm}  
M^{vv'}_{75} = - \d_{v',v} - \d_{v',v+\hat y} ,  \quad  M^{vv'}_{76} = - \d_{v',v} - \d_{v',v+\hat x} , \quad
 M^{vv'}_{77} = 2 \, \d_{v',v}. \nonumber
\neqa}}
The remaining non zero entries can be found following the pattern given above.

To compare these results to the continuum Einstein--Hilbert action it is convenient to change 
variables from the $\d_e$ labeling the fluctuations of the edge lengths to 
(suitable discretizations of) the metric perturbations 
$h_{\mu\nu} \equiv g_{\mu\nu}-g_{\mu\nu}^{\rm bg}$. The latter can be discretized 
using the lattice cartesian coordinates $\hat x^\mu=(1,0,0)$, etc., introduced in section \ref{Sec4}.
The change of variables can be found expanding to first order in the $\delta$ variables the following relations,
\eqa\label{trafo1}
&& \hat g_{\mu\nu} \, \hat x^\mu \, \hat x^\nu  = (l_{1}+\delta_{1})^2  \quad, \ldots , \no
&& g_{\mu\nu} \, (\hat x+\hat y)^\mu \, (\hat x+\hat y)^\nu = (l_{4}+\delta_{4})^2 \quad, \ldots, \no
&& g_{\mu\nu} (\hat x+\hat y+\hat z)^\mu \,(\hat x+\hat y+\hat z)^\nu = (l_{7}+\delta_{7})^2. \nonumber 
\neqa
These relations parametrize the six metric components per point in terms of seven edge lengths per lattice vertex.
This brings us immediately to the problem of the spurious variables: one of the seven $\d_{\hat e}$ is redundant.
In principle we can define a degenerate transformation $\d_{\hat e}^v\mapsto h_{\mu\nu}(x)$, but it is more convenient
to introduce an auxiliary seventh metric variable per vertex, $h_7^v$. 
We can then introduce discrete metric variables $h_{\hat e}^v \mapsto (h_{\mu\nu}(x), h_7^v)$, where for the
first six variables we have the mapping $x\mapsto v$ and $(11)(22)(33)(12)(13)(23)\mapsto 123456$.
Because of the absence in \Ref{QvApp} of correlations
between edges starting at $v$ and edges starting at $v+\hat x +\hat y+ \hat z$
(this can be seen from the lack of terms like $\d_{v', v+\hat x +\hat y+ \hat z}$ in \Ref{QvApp})
it is convenient to relate this spurious variable to the fluctuation $\d_7$, 
requiring that $\d_1 \ldots \d_6$ are functions of the six metric variables only.
This can be achieved with the following transformation,
\equ\label{Ttrasf}
h^v_{\hat e} =\sum_{v',\hat e'} T^{vv'}_{\hat e \hat e'} \delta^{v'}_{\hat e'}
\nequ
with
\eqa\label{T}
 && T^{vv'} = 
\left( {\scriptsize \begin{array}{ccc}
2a \, \d_{v',v} & 0 & 0 \\  0 & 2b \d_{v',v} &0 \\  0&0& 2c \d_{v',v} \\ -a \d_{v',v} & -b \d_{v',v} &0 \\
-a \d_{v',v} & 0 & -c \d_{v',v} \\ 0& -b \d_{v',v} & -c \d_{v',v} \\  \f a2 (\d_{v',v+\hat y}+\d_{v',v+\hat z})& 
\f b2(\d_{v',v+\hat x}+\d_{v',v+\hat z}) & \f c2(\d_{v',v+\hat x}+\d_{v',v+\hat y}) 
\end{array} }\right.
\\ \no \no \nonumber && \hspace{-1.6cm}
\left.  {\scriptsize \begin{array}{cccc}
0&0&0&0 \\ 0&0&0&0 \\ 0&0&0&0 \\ \sqrt{a^2+b^2} \, \d_{v',v}&0&0&0 \\ 
0&\sqrt{a^2+c^2} \,\d_{v',v}&0&0 \\ 0&0&\sqrt{b^2+c^2} \,\d_{v',v}&0 \\
-\f12{\sqrt{a^2+b^2}}(\d_{v',v}+\d_{v',v+\hat z}) & -\f12{\sqrt{a^2+c^2}}(\d_{v',v}+\d_{v',v+\hat y}) 
&  -\f12{\sqrt{b^2+c^2}}(\d_{v',v}+\d_{v',v+\hat x}) & {\sqrt{a^2+b^2+c^2}}\,\d_{v',v}
\end{array} } \right).
\neqa
Notice that $T^{vv'}$ has length dimension one, and thus $h_{\hat e}^v$ length dimension two.
This change of variables leads to a decoupling of the $h_7^v$ variables:
indeed, in these new variables the action reads 
\equ
S[h^v_{\hat e}]= \f12 \sum_{{v, v'} \atop {1\leq \hat e,{\hat e'}\leq 6}} 
h^v_{\hat e} \, L^{vv'}_{\hat e{\hat e'}}\, h^{v'}_{\hat e'} -\f1{abc}\sum_{v} (h^v_7)^2.
\nequ
Remarkably, it can be proved that the first summand gives the discretization of the second variation of the Einstein--Hilbert action $\sqrt{g}R$ around the background metric $g^{\rm bg}=\text{diag}(a^2,b^2,c^2)$,
\equ\label{S2}
S^{(2)}[h_{\mu\nu}; g^{\rm bg}] = \sqrt{g^{\rm bg}}
\left(
-\tfrac{1}{4}h_{\mu\nu}\partial_\sigma\partial^\sigma h^{\mu\nu}+
\tfrac{1}{2}h_{\mu\nu}\partial_\sigma \partial^\nu h^{\mu\sigma }-
\tfrac{1}{2}h_{\mu\nu}\partial^\mu\partial^\nu h_\sigma^\sigma+
\tfrac{1}{4}h^\mu_\mu \partial^\nu \partial_\nu h^\sigma_\sigma 
\right).
\nequ
Furthermore, an explicit calculation shows that
\equ\label{scaling}
S^{(2)}[h_{\mu\nu}; g^{\rm bg}] \equiv \f1{abc} \, S^{(2)}[h_{\mu\nu};
\eta].
\nequ
where $\eta=\text{diag}(1,1,1)$ is the euclidean metric. This somewhat surprising result shows
that the simple scaling of \Ref{defM} (or \Ref{QvApp}) is also a property of the continuum action 
(the additional lengths scalings present in (\ref{defM}) and \Ref{QvApp} 
are absorbed into the transformation \Ref{Ttrasf} from the $\d$-variables to the metric variables). 
We stress that this is peculiar to three dimensions: the same action \Ref{S2} in 4d
does not have the property \Ref{scaling}. As we comment below, 
this 3d scaling \Ref{scaling} is related to the fact that three dimensional gravity has no local degrees of freedom.

The remaining step is the analysis of the gauge symmetries, and their fixing.
In terms of the $\d$-variables, the total lattice kinetic term (\ref{QvApp}) (and so \Ref{defM} in the main text) 
displays three null vectors per vertex, given by 
\eqa\label{traslnull}
l_{\hat e} \, w^{\hat x}_{\hat e} &=& 
(\delta_{v',v+\hat x}-\delta_{v',v},0,0,\delta_{v',v+\hat x+\hat y}-\delta_{v',v}, 
\delta_{v',v+\hat x+\hat z}-\delta_{v',v},0,\delta_{v',v+\hat x+\hat y+\hat z}-\delta_{v',v}),  \no
l_{\hat e} \, w^{\hat y}_{\hat e} &=& 
(0,\delta_{v',v+\hat y}-\delta_{v',v},0,\delta_{v',v+\hat x+\hat y}-\delta_{v',v},0,
\delta_{v',v+\hat y+\hat z}-\delta_{v',v},\delta_{v',v+\hat x+\hat y+\hat z}-\delta_{v',v}),   \no
l_{\hat e} \, w^{\hat z}_{\hat e} &=& (0,0,\delta_{v',v+\hat z}-\delta_{v',v},
0,\delta_{v',v+\hat x+\hat z}-\delta_{v',v},\delta_{v',v+\hat y+\hat z}-\delta_{v',v},
\delta_{v',v+\hat x+\hat y+\hat z}-\delta_{v',v}).
\neqa
These null vectors correspond to the induced change of edge length if a vertex is translated in either the $\hat x, \hat y$ or $\hat z$ direction. Therefore \Ref{defM} has a gauge symmetry given by translations in 3d space,
which is what remains of the continuum diffeomorphism symmetry on the regular lattice.

Understanding these symmetries in terms of their continuum counterpart helps choosing a gauge
in which the continuum limit of the graviton propagator looks more transparent. We know from the continuum calculations that the propagator is simplest in the harmonic gauge
$C_\mu(x) \equiv \partial^\rho h_{\mu\rho }-\f12 \partial_\mu h^\rho{}_\rho = 0$.
This quantity can be immediately discretized using the correspondence with
the $h_{\hat e}^v$ variables introduced above, and one obtains 
$C_\mu(x) \mapsto \sum_{\hat e} C_{\mu \hat e}^v \, h_{\hat e}^v$ where
\eqa\label{C}
C_{\mu \hat e}^v = 
 \left( {\small { \begin{array}{ccc}
\f1{2 a^{2}}(-\delta_{v',v} + \delta_{v',v+\hat x}) & 
\f1{2 b^2}(\delta_{v',v}-\delta_{v',v+\hat x}) &
\f1{2 c^2}(\delta_{v',v}-\delta_{v',v+\hat x}) \\
\f1{2 a^2}(\delta_{v',v}-\delta_{v',v+\hat y}) & \f1{2 b^2}(-\delta_{v',v} + \delta_{v',v+\hat y }) &
\f1{2 c^2}(\delta_{v',v}-\delta_{v',v+\hat y}) \\
\f1{2 a^2} (\delta_{v',v}-\delta_{v',v+\hat z}) & \f1{2 b^2}(\delta_{v',v}-\delta_{v',v+\hat z}) &
\f1{2 c^2}(-\delta_{v',v} + \delta_{v',v+\hat z }) 
\end{array} }}\right. \no\no
\left. {\small { \begin{array}{ccc}
\f1{b^2}(\delta_{v',v}-\delta_{v',v-\hat y}) &
\f1{c^2}(\delta_{v',v}-\delta_{v',v-\hat z}) & 0 \\
\f1{a^2}(\delta_{v',v}-\delta_{v',v-\hat x}) &
0 & \f1{c^2}(\delta_{v',v}-\delta_{v',v-\hat z}) \\
0 & \f1{a^2}(\delta_{v',v}-\delta_{v',v-\hat x}) &  \f1{b^2}(\delta_{v',v}-\delta_{v',v-\hat y})
\end{array} }} \right) .
\neqa
Analogously, one can discretize the gauge fixing term $\f12 \sqrt{g^{\rm bg}}  \, C^\mu C_\mu$
leading to \Ref{finito} in the main text. This completes the derivation of the main results
presented in section \ref{Sec4}.

To conclude, notice the non trivial scaling of \Ref{C}, and in particular that it does not agree with the simple scaling of (\ref{scaling}). This non trivial scaling is crucial in order to obtain the correct scaling of
the discretized laplacian in \Ref{baction5},
\equ\label{lapl}
\Delta = \sum_{{\hat \mu}} 
\big( \d_{v',v+{\hat \mu}}+ \d_{v',v-{\hat \mu}} - 2 \d_{v',v}\big)/{l_\mu^2}.
\nequ

The fact that the non trivial scaling of \Ref{lapl} comes from the gauge fixing 
term\footnote{Indeed if one added a term $C_{\mu}C_\nu\eta^{\mu \nu}/abc$ 
which has the same scaling of (\ref{scaling})
one would end up with the laplacian for the background metric
$\eta=\text{diag}(1,1,1) $, namely 
\equ\label{Wwrong}
\Delta = \sum_{\hat \mu} 
\big( \d_{v',v+\hat \mu}+ \d_{v',v-\hat \mu} - 2 \d_{v',v}\big). 
\nequ 
Notice that of course $C_{\mu}C_\nu\eta^{\mu \nu}/abc$ is not covariant 
(it has the wrong background metric) thus it is not an admissable gauge fixing term.} 
and not from the kinetic term of the action might look confusing at first. However, this is a well known feature of
3d general relativity: it is a theory with no
local degrees of freedom, and the propagator we are computing
is a pure gauge quantity.\footnote{In particular, notice that as the second variation of the
Einstein Hilbert action (\ref{S2}) is given by a quadratic form of rank $3$
we could also use (minus) this term as a gauge fixing, which would lead to
a vanishing graviton, corresponding to no propagation at all. This of course is not regarded as a proper gauge
fixing, because the quadratic form defining the gauge fixed action has now
minimal rank instead of maximal rank.} Consistently, its dynamical
properties are completely determined by the gauge.
In 4 dimensions on the other hand \Ref{scaling} is no longer valid, hence
the above mechanism does not apply.

\end{document}